\documentclass[
    aps,prx,reprint,            
    superscriptaddress,         
    amsfonts,amsmath,amssymb,   
]{revtex4-2}

\usepackage{braket} 
\usepackage{graphicx}
\usepackage{xspace}
\usepackage{xcolor}
\usepackage{siunitx}
\usepackage{multirow}
\usepackage{verbatim}
\usepackage[colorlinks=true,citecolor=blue,urlcolor=blue,linktocpage=true,linkcolor=blue]{hyperref}
\usepackage[capitalize]{cleveref}

\graphicspath{{./graphics/}}

\DeclareSIUnit{\cps}{\per\s} 
\DeclareSIUnit{\afu}{\hertz} 
\newcommand*{\DM}{\ensuremath{\mathrm{DM}}\xspace} 
\newcommand*{\MPl}{\ensuremath{M_\mathrm{Pl}}\xspace} 
\newcommand*{\Ncopy}{\ensuremath{\mathcal{N}}\xspace} 
\newcommand*{\Th}{\texorpdfstring{\textsuperscript{229}T\MakeLowercase{h}\xspace}{Th-299\xspace}} 
\newcommand*{\f}{\ensuremath{\nu}\xspace} 
\newcommand*{\fTh}{\ensuremath{\f_{\text{Th}}}\xspace} 
\newcommand*{\te}{\ensuremath{t_e}\xspace} 
\newcommand*{\td}{\ensuremath{t_d}\xspace} 
\newcommand*{\dN}{{\ensuremath{\Delta N}\xspace}} 

\begin{document}

%
%

\title{Searching for dark matter with the \Th nuclear lineshape from laser spectroscopy}

\author{Elina~Fuchs}
\email{elina.fuchs@itp.uni-hannover.de}
\affiliation{Physikalisch-Technische Bundesanstalt, Bundesallee 100, Braunschweig, 38116, Germany}
\affiliation{Institut f\"{u}r Theoretische Physik, Leibniz Universit\"{a}t Hannover, Appelstraße 2, Hannover, 30167, Germany}
\author{Fiona~Kirk}
\email{fiona.kirk@itp.uni-hannover.de}
\affiliation{Physikalisch-Technische Bundesanstalt, Bundesallee 100, Braunschweig, 38116, Germany}
\affiliation{Institut f\"{u}r Theoretische Physik, Leibniz Universit\"{a}t Hannover, Appelstraße 2, Hannover, 30167, Germany}
\author{Eric~Madge}
\email{eric.madgepimentel@uam.es}
\affiliation{Department of Particle Physics and Astrophysics, Weizmann Institute of Science, Rehovot 761001, Israel}
\affiliation{Instituto de F\'isica Te\'orica UAM-CSIC {\normalfont and} Departamento de F\'isica Te\'orica, Universidad Aut\'onoma de Madrid, C/~Nicol\'as Cabrera 13--15, Cantoblanco, Madrid 28049, Spain}
\author{Chaitanya~Paranjape}
\email{chaitanya.paranjape@weizmann.ac.il}
\affiliation{Department of Particle Physics and Astrophysics, Weizmann Institute of Science, Rehovot 761001, Israel}
\author{Ekkehard~Peik}
\email{ekkehard.peik@ptb.de}
\affiliation{Physikalisch-Technische Bundesanstalt, Bundesallee 100, Braunschweig, 38116, Germany}
\author{Gilad~Perez}
\email{gilad.perez@weizmann.ac.il}
\affiliation{Department of Particle Physics and Astrophysics, Weizmann Institute of Science, Rehovot 761001, Israel}
\author{Wolfram~Ratzinger} 
\email{wolfram.ratzinger@weizmann.ac.il}
\affiliation{Department of Particle Physics and Astrophysics, Weizmann Institute of Science, Rehovot 761001, Israel}
\author{Johannes~Tiedau}
\email{johannes.tiedau@ptb.de}
\affiliation{Physikalisch-Technische Bundesanstalt, Bundesallee 100, Braunschweig, 38116, Germany}

\date{\today}

\begin{abstract}
The recent laser excitation of the low-lying \Th isomer transition is starting a revolution in ultralight dark matter searches. The enhanced sensitivity of this transition to the large class of dark matter models dominantly coupling to quarks and gluons will ultimately allow us to probe coupling strengths eight orders of magnitude smaller than the current bounds from optical atomic clocks, which are mainly sensitive to dark matter couplings to electrons and photons. 
We argue that, with increasing precision, observations of the \Th excitation spectrum will soon give world-leading constraints. 
Using data from the  pioneering laser excitation of \Th 
by~\citeauthor{Tiedau:2024obk} [\href{https://doi.org/10.1103/PhysRevLett.132.182501}{Phys.~Rev.~Lett.~{\bf132}, 182501 (2024)}], we present a first dark matter search in the excitation spectrum.
While the exclusion limits of our detailed study of the lineshape are still below the sensitivity of currently operating clock experiments, we project the measurement of~\citeauthor{Zhang:2024ngu} [\href{https://doi.org/10.1038/s41586-024-07839-6}{Nature~{\bf 663}, 63 (2024)}] to surpass it. 
\end{abstract}

\maketitle


%
\section{Introduction}
\label{sec:intro}
%

Although abundant evidence from astrophysical and cosmological observations via gravity supports the existence of dark matter~(DM), 
little is known about its fundamental properties.
Theories of ultralight dark matter~(ULDM) bosons (scalar or pseudo-scalar) provide us with one of the simplest frameworks for DM. 
Well-motivated models of ULDM include the axion~\cite{Preskill:1982cy,Abbott:1982af,Dine:1982ah,Hook:2018dlk,DiLuzio:2020wdo} of quantum chromodynamics~(QCD), the dilaton~\cite{Cho:1998js,Damour:2010rm,Damour:2010rp,Arvanitaki:2014faa} (though see Ref.~\cite{Hubisz:2024hyz}), the relaxion~\cite{Graham:2015ifn,Banerjee:2018xmn,Chatrchyan:2022dpy}, and possibly other forms of Higgs-portal models~\cite{Piazza:2010ye}. Finally, it was recently shown that the Nelson-Barr framework~\cite{Nelson:1983zb,Barr:1984fh,Barr:1984qx} that also addresses the strong-CP problem, leads to a viable ULDM candidate~\cite{Dine:2024bxv}.
All of these models predict the ULDM to couple dominantly to the Standard Model~(SM) QCD sector, i.e.\ the quarks and the gluons, leading to oscillations of nuclear parameters~\cite{Flambaum:2006ip,Damour:2010rm,Damour:2010rp,Kim:2022ype}.
In addition, one can consider signals associated with transient phenomena such as topological defects~\cite{Derevianko:2013oaa}.

Variations of SM parameters can be searched for by comparing the rates of two frequency standards that exhibit different dependencies on the parameters in question~\cite{Uzan:2010pm,Damour:2010rp,Arvanitaki:2014faa,Stadnik:2015kia}.
Laboratory limits on these variations have been obtained from various clock-comparison experiments based on atomic or molecular spectroscopy as well as cavities and mechanical oscillators (see Refs.~\cite{Uzan:2010pm,Safronova:2017xyt,Antypas:2020rtg,Antypas:2022asj} for a review).
However, these frequency standards mostly rely on electronic properties, whereas their sensitivity to changes in the nuclear sector is largely suppressed. 
In optical clocks, nuclear properties enter via the hyperfine structure and the reduced mass, but their relative contributions to the transition frequency is typically only of order $10^{-6}$ and $10^{-5}$ respectively. 
The contribution from an oscillation of the charge radius is around $10^{-3}$~\cite{Banerjee:2023bjc}.
Rotational and vibrational transitions in molecules give an order $\numrange{e-2}{e-1}$ contribution~\cite{Oswald:2021vtc}. 
Clocks based on hyperfine transitions and mechanical oscillators (Cs~clock, hydrogen maser, quartz oscillator)~\cite{Hees:2016gop,Kennedy:2020bac,Kobayashi:2022vsf,Sherrill:2023zah,Zhang:2022ewz,Campbell:2020fvq}, exhibit a dependence on nuclear magnetic moments or masses already at leading order, however, these do not reach the accuracy and stability of optical clocks.
Future optical clocks based on highly charged ions~\cite{Kozlov:2018mbp}, as well as pure rotational-vibrational transitions in molecular clocks with nearly degenerate energy levels~\cite{Kozyryev:2018pcp} or large overtone transitions~\cite{Madge:2024aot}, can enhance the sensitivity by up to two or three orders of magnitude.

In contrast, the cancellation of electromagnetic and strong contributions resulting in the unusually low-lying isomer transition of \Th~\cite{Flambaum:2008ij,Berengut:2009zz} promises a sensitivity enhanced by $\mathcal{O}(10^8)$ relative to existing probes of QCD~\cite{Arvanitaki:2014faa,Kim:2022ype}.
This enormous leap in sensitivity to the strong sector, and consequently to ULDM models coupling to it, implies that unexplored parameter space can be tested even before a nuclear clock becomes available.

In Ref.~\cite{Tiedau:2024obk}, for the first time, the isomer state was resonantly excited by a tabletop tunable vacuum-ultraviolet~(VUV) laser system~\cite{Thielking2023_VUVlaser_NJP} using \Th dopant ions in a crystal, as discussed in Refs.~\cite{PEIK_2009,Rellergert:2010,Kazakov_2012}. The host crystal in this experiment was CaF$_2$~\cite{Beeks:2022dnl}. The relative uncertainty on the observed resonance was decreased to $\mathcal{O}(10^{-6})$, which corresponds to an improvement of almost three orders of magnitude with respect to the measurement of the transition energy via the detection of the radiative decay~\cite{Kraemer:2022gpi}. The excitation energy observed by Ref.~\cite{Tiedau:2024obk} was confirmed by Ref.~\cite{Elwell:2024qyh} with a different host crystal, LiSrAlF$_6$. Another six orders-of-magnitude improvement in terms of accuracy on the transition frequency was obtained in laser excitation with single modes of a VUV frequency comb~\cite{Zhang:2024ngu}.

\begin{figure}
    \centering
    \includegraphics[width=\columnwidth]{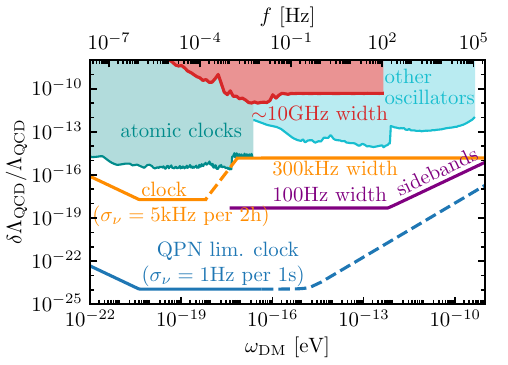}
    \caption{%
        Searches for periodic variations of the QCD scale~$\Lambda_\mathrm{QCD}$ over time. 
        The bound derived in this work by performing a lineshape analysis of the spectrum recorded in Ref.~\cite{Tiedau:2024obk} is shown in red, whereas the orange line shows the projection based on the results of Ref.~\cite{Zhang:2024ngu}. 
        The purple line indicates the prospective reach of a lineshape analysis limited by magnetic fields in the host crystal. For comparison, current bounds from atomic clocks~\cite{Banerjee:2023bjc,Hees:2016gop,Kennedy:2020bac,Kobayashi:2022vsf,Sherrill:2023zah} and other oscillators~\cite{Oswald:2021vtc,Campbell:2020fvq,Zhang:2022ewz} are shown in dark and light cyan respectively, whereas a projection for a future quantum projection noise~(QPN)-limited single-ion nuclear clock~\cite{Peik:2020cwm,Arvanitaki:2014faa,Banerjee:2020kww} is indicated by the blue line. 
        For details, see discussion in \cref{sec:LambdaQCDbounds}.
    }
    \label{fig:variation_QCD_scale}
\end{figure}

In this work we show, for the first time, how lineshape measurements of the low-lying isomer transition of \Th can be exploited for new physics searches, without requiring a fully developed nuclear clock.
We demonstrate this procedure in detail on the data of Ref.~\cite{Tiedau:2024obk}.
The implications of the enhanced sensitivity, as well as the rapidly decreasing widths and growing accuracy, can most clearly be seen from \cref{fig:variation_QCD_scale}, where we compare searches for periodic variations of the QCD scale~$\Lambda_\mathrm{QCD}$. 
In parts of the parameter space, the lineshape bound derived in this work, shown in red, is only one to three orders of magnitude less stringent than the current clock bounds~(cyan).
It should be straightforward to analyse the results of Ref.~\cite{Zhang:2024ngu} in the same manner. The corresponding estimates are shown in orange.
Since the limiting factor for our method is the observed width of the transition, which was so far dominated by the laser linewidth, the observation of Ref.~\cite{Zhang:2024ngu} with a full-width at half-maximum reduced from \SI{20}{\GHz} to \SI{300}{\kHz}, i.e.\ by $\mathcal{O}(\num{e-5})$ compared to Ref.~\cite{Tiedau:2024obk}, highlights the potential of our approach.
Eventually, the linewidth will no longer be limited by the laser width but by the effects of the host crystal or of the relativistic Doppler effect. Assuming a linewidth of $\sim\SI{100}{\Hz}$~\cite{Kazakov_2012} we find the projections shown in purple.
While this article focuses on variations of the QCD scale, the results can easily be reinterpreted in terms of variations in the fine-structure constant or quark masses, or in terms of a combination of these effects. 

We note that this method of analyzing the lineshape can be applied to many systems where clock operation (with a laser whose frequency is locked to the center of the resonance line) has not 
yet been achieved. An example for such a system is the recent X-ray laser excitation of the $^{45}$Sc isomer \cite{Shvydko:2023} and future experiments with trapped \Th ions.

This article is organised as follows. We start with a short review of the enhanced sensitivity of the \Th isomer transition to ULDM in \cref{sec:Th_sensitivity}.
We then derive limits on periodic modulations of the \Th frequency in \cref{sec:lineshape_bounds}.
After a brief introduction to the experimental setup for the laser excitation in \cref{sec:Th_excitation}, \cref{sec:bounds_estimate} gives some intuition  as to how new physics could manifest itself in the measured lineshape. We corroborate our analytic estimates by performing a numerical analysis in \cref{sec:lineshape_discussion}.
Further details on the analysis can be found in \cref{sec:sidebands,sec:MCMCandODR}.
The lineshape bounds are then used to place limits on oscillations of the QCD scale~$\Lambda_\mathrm{QCD}$ in \cref{sec:LambdaQCDbounds}.
In \cref{sec:models} we define new physics models that can be probed using nuclear lineshape data and recast the experimental bounds derived in \cref{sec:lineshape_bounds} for these more specific setups.
We conclude with a short summary of this work in \cref{sec:conclusion}.

%
\section{Sensitivity of \Th to ULDM}
\label{sec:Th_sensitivity}
%

Ultralight dark matter~(ULDM) refers to bosonic DM candidates with masses below $m_\DM \lesssim \SI{1}{\eV}$. Assuming that ULDM accounts for the entire DM abundance, its occupation number is at least one particle per de~Broglie volume. Consequently, 
ULDM can be described as a classical field, oscillating at a frequency given by its mass~$m_\DM$ and with a coherence time of $\tau_\mathrm{coh} \sim 1/(m_\DM v_\DM^2)$, where $v_\DM \sim \num{e-3}$ is the DM velocity dispersion, assuming it is comparable to the velocity of the solar system within the Milky Way galaxy.
If the DM field is a scalar or a pseudoscalar, it can be described by 
\begin{align}
    \phi(t,\mathbf{x}) \simeq \phi_0 \cos\left(m_\DM t + \varphi_\DM \right)\,,
\end{align}
where the amplitude~$\phi_0$ is related to the DM density $\rho_\phi=\frac{1}{2}m_\DM^2 \phi_0^2$, and $\varphi_\DM$ is a random phase. In the above expression, we have neglected spatial dependencies that are small due to the non-relativistic nature of the field.

The interactions of ULDM with the SM can induce changes in what would otherwise be fundamental constants of nature, namely the electric and strong interaction strength as well as the fermion masses.
These take the form
\begin{align}
    \frac{\delta X}{X} \propto \cos\left(\omega_\DM t + \varphi_\DM\right)\,,
\end{align}
where $X$ is a fundamental constant such as the QCD scale $\Lambda_\mathrm{QCD}$, the fine-structure constant~$\alpha_\mathrm{EM}$, or the electron and light quark masses~$m_e$ and~$m_q$, $q=u,d,s$.
We consider concrete examples of couplings leading to such oscillations in \cref{sec:models}.
The frequency of oscillation $\omega_\DM$ is given by the DM mass $m_\DM$ or $2\, m_\DM$, depending on whether one considers a linear or quadratic coupling.

As a consequence of the time-varying fundamental constants, observables like atomic and nuclear transition energies change as well.
In particular, the \Th transition frequency takes the form
\begin{equation}
    \f(t)\simeq\f_0+\delta\f_\DM \cos(\omega_\DM t+\varphi_\DM)\,. \label{eq:oscillating freq}
\end{equation}
Above, $\f_0$ denotes the nuclear transition frequency in the absence of DM, and $\delta\f_\DM$ is the amplitude of the DM-induced variation. 

The amplitude of the oscillations of the nuclear transition frequency generally results from a combination of the DM coupling and DM density, as well as the sensitivity of the transition to the DM couplings in question. 
One can parameterize the relative time-variation of the transition frequency as
\begin{align}
    \frac{\delta \f_\DM}{\f_0} = \hspace{-1em} \sum_{\substack{X\in\{\alpha_\mathrm{EM},m_{e},\\\hspace{1.5em}\Lambda_\mathrm{QCD},m_{q}\}}} \hspace{-1em} K_{X}\,\frac{\delta X}{X},\label{eq:delX}
\end{align}
where $K_{X}$ denotes the sensitivity coefficients of the transition $\f(t)$ to a variation of the SM parameter $X$. 

The outstanding feature of the \Th isomer transition is its potentially enhanced sensitivity to $\Lambda_\mathrm{QCD}$, $(m_u+m_d)$ and $\alpha_\mathrm{EM}$ through what appears to be a fine-tuning of nature.
The enhanced sensitivity to $\alpha_\mathrm{EM}$ can be understood by splitting the isomeric transition energy into the contributions of the Coulomb energy $\Delta E_\mathrm{C}$ and
of the QCD binding energy $\Delta E_\mathrm{nuc}$,  
\begin{equation}
    \fTh = \Delta E_\mathrm{nuc} - \Delta E_\mathrm{C} \approx \SI{8}{eV}\,.
\end{equation}
The Coulomb contribution to the binding energy is proportional to $\alpha_\mathrm{EM}$.
One therefore finds
\begin{equation}
    K_{\alpha_\mathrm{EM}} \sim \frac{\mathrm{d} \log\fTh}{\mathrm{d} \log\alpha_\mathrm{EM}}\sim\frac{-\Delta E_\mathrm{C}}{\fTh}\,,
\end{equation}
and analogous expressions are expected to hold for the other parameters. In principle, these sensitivity parameters are non-perturbative, however, in the literature one can find recent attempts to estimate $K_{\alpha_\mathrm{EM}}$ using a classical geometrical approach~\cite{Flambaum:2006ak,Flambaum:2008ij,Berengut:2009zz,Berengut:2010zj,Beeks:2024xnc}. 
This approach is limited both because it neglects any quantum mechanical effects responsible for the existence of the isomer excitation, and because it does not include other known contributions to the Coulomb energy. 
Further, even at the classical level, the transition energy difference depends on a host of parameters that are subject to large experimental uncertainties~\cite{Caputo:2024doz}. 
One can extend this approach or consider an effective-field-theory-inspired quantum model of \Th to further deepen our knowledge of $K_{\alpha_{\mathrm{EM}}}$, as well as its relation to $K_{\Lambda_\mathrm{QCD}}$~\cite{Caputo:2024doz}, which is more relevant to this work. Very roughly, we expect 
\begin{equation}
\left|K_{\Lambda_\mathrm{QCD}}\right|\!\sim \left|K_{\alpha_\mathrm{EM}}\right|\! \sim \SIrange{0.1}{1}{\MeV}/\fTh\sim \numrange{e4}{e5}\!.\label{eq:Ks}
\end{equation} 

The sensitivity gain of the nuclear clock is particularly prominent in the case of DM that only couples directly to the strong/nuclear sector.
As briefly described in the introduction, current atomic clock comparisons of electronic transitions have a reduced sensitivity $K_{\Lambda_\mathrm{QCD}}^{\text{atom}}\sim\mathcal{O}(\numrange{e-3}{e-2})$ to nuclear parameters as the corresponding contribution to the transition frequencies is subdominant (e.g.\ Refs.~\cite{Hees:2016gop,Banerjee:2023bjc}), whereas the accuracy of the frequency measurement in comparisons involving hyperfine transitions or quartz oscillators cannot compete with the current best optical clocks, $\left[{\delta \f}/{\f_0}\right]_\mathrm{atom} \sim \mathcal{O}\left(\numrange{e-5}{e-2}\right) \times \left[{\delta \f}/{\f_0}\right]_\mathrm{best}$ (e.g.\ Refs.~\cite{Oswald:2021vtc,Campbell:2020fvq,Kobayashi:2022vsf,Zhang:2022ewz,Sherrill:2023zah,Kennedy:2020bac} compared to $\delta \f/\f \sim 10^{-18}$ from Ref.~\cite{BACON:2020ubh}). 
The realisation of a state-of-the-art nuclear clock based on the isomer transition and limited by quantum projection noise would therefore enhance the sensitivity to a variation of the nuclear scales by a factor of 
\begin{align}
    \frac{\left(\!\frac{\delta\Lambda_\mathrm{QCD}}{\Lambda_\mathrm{QCD}}\!\right)_{\!\mathrm{atom}}}{\left(\!\frac{\delta\Lambda_\mathrm{QCD}}{\Lambda_\mathrm{QCD}}\!\right)_{\!\mathrm{nuc}}} = \frac{\left(\!K_{\Lambda_\mathrm{QCD}}^{-1} \!\frac{\delta \f}{\f_0}\!\right)_{\!\mathrm{atom}}}{\left(\!K_{\Lambda_\mathrm{QCD}}^{-1} \frac{\delta\f}{\f_0}\!\right)_{\!\mathrm{nuc}}}
    \sim \mathcal{O}(\numrange{e8}{e10})\,.
\end{align}
For simplicity, and to demonstrate the DM search potential of interrogating the \Th isomer transition, we concentrate on a nucleophilic scalar field coupling to~$\Lambda_\mathrm{QCD}$. In principle, the comparison of two frequencies probes the difference between the respective sensitivity coefficients, however, in our case the sensitivity of the nuclear clock vastly dominates and we can neglect the impact of DM on the second frequency.

%
\section{Bounds on ULDM from lineshape}
\label{sec:lineshape_bounds}
%

The use of a narrowband VUV laser and improved control of systematic frequency shifts will eventually allow us to build a highly stable and accurate nuclear clock. 
On the way there, it is possible to probe models of new physics via the analysis of variations in the lineshape that they may cause. 
Here, we describe such a search focusing on models of ULDM coupled to nuclei.

\subsection{Laser excitation of \Th}
\label{sec:Th_excitation}

Schematically, the laser excitation of the \Th isomer presented in Refs.~\cite{Tiedau:2024obk,Elwell:2024qyh,Zhang:2024ngu} is achieved as follows:
A \Th-doped crystal is irradiated with light from a tunable VUV laser for a time period~\te.
Subsequently, VUV fluorescence photons from isomer decays are detected in a photon multiplier tube~(PMT) and the PMT counts are recorded for a period~\td.
After each excitation time~\te the laser is turned off to avoid scattering light on the PMT. The fluorescence light from the crystal during the detection time~\td is focused and filtered with two dielectric mirrors to minimize the radioluminescence signal. Finally, the signal from the PMT is recorded with a threshold counting card.
This excitation and measurement sequence is then repeated, varying the frequency of the VUV laser and thereby recording the excitation spectrum of the \Th nuclear resonance curve.

In the following, we mainly focus on the frequency scans of the highly-doped CaF\textsubscript{2} crystal~X2 shown in Fig.~2 of Ref.~\cite{Tiedau:2024obk}.
The data is taken with an excitation time of $\te = \SI{120}{\s}$ and a measurement time of $\td = \SI{150}{\s}$.
The resonance is scanned twice, once with decreasing and once with increasing laser frequency. The time difference between the beginning of the two scans is $T=\SI{130}{\minute}$.

Since the time $t_e+t_d$ between two subsequent measurement points is smaller than the fluorescence lifetime $\tau=\SI{630}{\s}$, the count rate also includes decays from isotopes excited during previous illumination cycles. The resulting resonance curves are asymmetric and depend on the scan direction.
Subtracting the decays from previous cycles, as well as the radioluminescence background counts from nuclear decays of \Th and its decay chain, produces symmetric resonance curves that coincide for the two scan directions.
The post-processed number of counts \dN\ from an excitation cycle starting at $t=0$ is then given by [cf.\ \cref{sec:MCMCandODR}]
\begin{align}
    \dN \propto 
    \int\limits_{0}^{\te}\!dt \int\!d\f\,L\big(\f;\f_L\big)\,I\big(\f;\fTh(t)\big)\, e^{-\frac{t-\te}{\tau}}\,,
\end{align}
where $\f_L$ and $\fTh$ are the frequencies of the laser and \Th resonance, and $L(\f;\f_L)$ and $I(\f;\fTh)$ are the corresponding lineshapes within the crystal.

\subsection{Analytic estimate}
\label{sec:bounds_estimate}

To gain some physical intuition as to how this variation in transition frequency manifests itself in the \Th excitation spectrum, it is helpful to consider a few interesting limits:
When the DM amplitude is much larger than the DM frequency, i.e.\ when $\delta\f_\DM \gg \omega_\DM/(2\pi)$, we can treat \cref{eq:oscillating freq} as the instantaneous transition energy at time $t$, such that the nucleus is only resonantly excited if the laser frequency matches $\f(t)$.
Since in Ref.~\cite{Tiedau:2024obk} the width of the laser is the dominant factor contributing to the width of the recorded resonance and hence limits the sensitivity to DM, all parameter space considered here lies within this regime.

Furthermore, if the period of the DM oscillation is much longer than the duration $T$ of the experiment, i.e. if $\omega_\DM T\ll2\pi$, we may approximate
\begin{equation}
    \f(t)\simeq\f_0+\delta\f_\DM \cos(\varphi_\DM) - \delta\f_\DM \sin(\varphi_\DM) \,\omega_\DM t\,.
\end{equation}
As the first and second terms are time-independent and hence indistinguishable, 
DM only manifests itself as a linear drift of the nuclear transition frequency. For example, if we do not observe a significant variation of the transition frequency~$\f$ between two measurements with uncertainty $\sigma_\mathrm{meas}$ and separated by a time $T$, we may marginalize over $\varphi_\DM$ ($|\sin(\varphi_\DM)|\to 2/\pi$) and constrain
\begin{equation}
    \delta\f_\DM \lesssim\frac{\pi}{2}\frac{\, \sqrt{2}\, \sigma_\mathrm{meas}}{\omega_\DM T}
    =   \frac{\pi\, \sigma_\mathrm{Th}}{\omega_\DM T}\,,
    \label{eq:limit_slow_adiabatic}
\end{equation}
at the $1\sigma$ level. Here we assumed the uncertainty $\sigma_\mathrm{Th}$ of the \Th frequency measurement, $\f_\mathrm{Th}\pm\sigma_\mathrm{Th}$,
to correspond to the uncertainty of the mean central frequency of the forward and backward scans and $ \sigma_\mathrm{meas} = \sqrt{2} \sigma_\mathrm{Th}$ to be the uncertainty on each frequency measurement ($\sigma_\mathrm{meas}=\sigma_{\nu(0)} = \sigma_{\nu(T)}$).

In the limit where the oscillation of the resonance frequency is much faster than all timescales relevant to the scanning process, $\omega_\DM T\gg2\pi$ but $\delta\f_\DM \gg \omega_\DM/(2\pi)$, the transition frequency takes on values between $\f_0\pm\delta\f_\DM$ within one scan cycle, leading to a broadening of the line by approximately $2\,\delta\f_\DM$\,. 
One is therefore able to obtain a $1\sigma$ constraint
\begin{equation}
    \delta\f_\DM \lesssim  \frac{\Delta \f}{2} \,,
    \label{eq:limit_fast_adiabatic}
\end{equation}
where $\Delta \f$ is the observed full-width at half-maximum of the line.

The spectral shape of this broadening effect can be estimated by modeling the spectral distribution of the DM-impacted transition line and averaging the distribution over a DM oscillation period $T_\DM=2\pi/\omega_\DM$, assuming $T_\DM$ is considerably shorter than the measurement time. If we further assume the nuclear lineshape to be faithfully described by a \mbox{$\delta$-distribution}, we obtain two contributions per frequency~\f and DM oscillation period, leading to
\begin{equation}\begin{aligned}
    \hspace*{-6pt}\langle I(\f)\rangle =\hspace{-4pt}\int\limits_0^{T_\DM}\hspace{-6pt}\frac{dt}{T_\DM}\, \delta(\f-\f(t)) 
    = \frac{\theta\!\left(1-\left|\frac{\f-\f_0} {\delta\f_\DM}\right|\right)/\pi}{\sqrt{\delta\f_\DM^2-\left({\f-\f_0}\right)^2}}\,,\hspace*{-3pt}
    \label{eq:intensity_adiabatic}
\end{aligned}\end{equation}
where $\theta$ is the Heaviside step function. In absence of DM, the observed lineshape is given by the convolution of the resonance lineshape without DM,
which is typically modelled by a Gaussian or a Lorentzian, with $I(\f)=\delta (\f - \f_0)$, whereas in presence of DM, the convolution of the resonance lineshape with \cref{eq:intensity_adiabatic} 
results in a double-resonance with two peaks at $\f_0\pm\delta\f_\DM$, i.e.\ the DM-modulation of the nuclear transition frequency splits the nuclear resonance into two.
This is a distinctive feature that can be searched for.
Since in Ref.~\cite{Tiedau:2024obk} only one peak is observed, the majority of the linewidth must be due to SM physics and we can place a bound on $\delta \f_\DM$ that is better than \cref{eq:limit_fast_adiabatic} by a factor of $\mathcal{O}(1)$.

As eluded to before, the analysis presented so far only applies for $\delta\f_\DM \gg \omega_\DM/(2\pi)$. Since we constrain $\delta\f \lesssim \Delta \f$, this is self-consistent up to DM masses corresponding to the laser width, so at the current stage up to roughly \SI{e-5}{\eV} (or \SI{1}{\GHz}). We do not expect competitive bounds for masses this large and therefore do not extend our analysis further. Since the resolution of the frequency is expected to improve rapidly, now that the nuclear transition can be excited in a controlled laboratory environment, we would like to mention that for $\delta\f_\DM \ll \omega_\DM/(2\pi)$ a search for sidebands separated by $\pm \omega_\DM/(2\pi)$ from the central frequency~$\f_0$ may be particularly fruitful. In \cref{sec:sidebands} we show how this phenomenon arises and how it relates to the broadening of the peak discussed above.

\subsection{Numerical analysis}
\label{sec:lineshape_discussion}

To corroborate our analytic estimate, we perform a numerical analysis of the data of Ref.~\cite{Tiedau:2024obk}. First the backgrounds and counts from previous excitation steps are subtracted. Then, the numbers of PMT counts $\dN_n$ in all steps $n$ of the excitation and detection cycle are fitted simultaneously.
Assuming a Lorentzian lineshape for the laser, the counts $\dN_n$ are given by
\begin{align}
    \dN_n = \dN_\mathrm{off} + \mathcal{N} \int\limits_0^1 \!dx\,\frac{e^{-x\,\te/\tau}}{1 + 4\left(\frac{\delta \f_n(x)}{\Delta_L}\right)^2}\,,
    \label{eq:pp_count_rate}
\end{align}
where $\Delta_L$ is the full laser linewidth at half-maximum, $\mathcal{N}$ is a constant prefactor, and $\delta\f_n(x)$ is the detuning between the laser frequency and the \Th resonance at time $x\, t_e$, $1 \geq x \geq 0$ during the $n$-th excitation period
($x=1$ and $x=0$ correspond to the beginning and the end of each excitation).
We include constant offsets $\dN_\mathrm{off}$ and $\delta\f_\mathrm{off}$ in the number of PMT counts and the detuning, respectively.
Further details are provided in \cref{sec:MCMCandODR}.

\begin{figure}
    \centering
    \includegraphics[width=\linewidth]{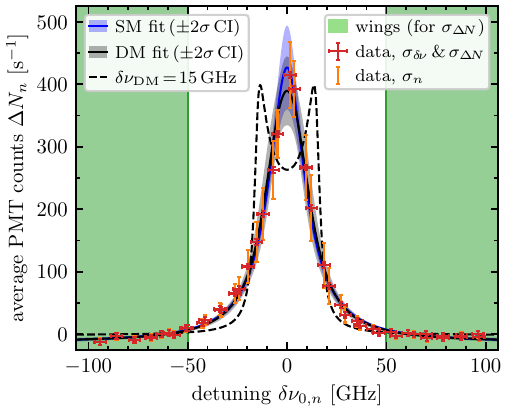}
    \caption{Lineshape fit using orthogonal distance regression to the experimental data with two-dimensional uncertainties. The uncertainties $\sigma_{\delta\f}$ on the frequency detuning take into account the triangular modulation of the laser frequency applied during the scan.
    The wings, shaded in green, correspond to the control region, in which the uncertainties $\sigma_{\dN}$ on the PMT counts were determined. 
    The orange error bars correspond to the uncertainties obtained by propagating $\sigma_{\delta\f}$ to the PMT count rate, as done in the MCMC fit.
    The best-fit parameters and corresponding standard deviations for the models without~(blue) and with~(black) DM with a characteristic frequency $\omega_\DM=\SI{1}{\afu}$ are shown in \cref{tab:bestfits}. The black dashed line illustrates the DM-dominated case (see also \cref{eq:intensity_adiabatic}) and was obtained by fixing $\delta \f_{\DM}=\SI{15}{\GHz}$.}
    \label{fig:Barad-Dur_vs_Minas-Tirith}
\end{figure}

A bound on $\delta\f_\DM$ can be obtained directly from a curve fit to the data presented in \cref{fig:Barad-Dur_vs_Minas-Tirith} in red. 
We perform the fit and the estimation of the respective uncertainties and exclusion bounds on the DM oscillation amplitude using 
two independent methods: Markov chain Monte Carlo~(MCMC) sampling and orthogonal distance regression~(ODR). In the second method, the probability distributions of the fitted parameters are approximated as being Gaussian.
The resulting mean parameter (for MCMC) or best-fit (for ODR) values and $1\sigma$ uncertainties for the fit without DM, and for a fit including DM with $\omega_\DM = \SI{1}{\afu}$, are listed in \cref{tab:bestfits}.
As expected, we observe that the fit is insensitive to the value of $\varphi_{\DM}$ in the limit where the period of the DM oscillation is much shorter than the detection time, i.e. $T_{\DM}\equiv 2\pi /\omega_{\DM} \ll t_d$. 

\begin{table}[t]
    \setlength{\tabcolsep}{1pt}
    \renewcommand{\arraystretch}{1.25}
    \begin{tabular}{@{\hskip 6pt}c@{\hskip 6pt}*{4}{rcl@{\hskip 6pt}}}
        \hline\hline
        & \multicolumn{6}{c}{SM fit} & \multicolumn{6}{c}{DM fit}\\
                                             & \multicolumn{3}{c}{MCMC} & \multicolumn{3}{c}{ODR} & \multicolumn{3}{c}{MCMC} & \multicolumn{3}{c}{ODR} \\\hline
        $\dN_{\mathrm{off}}$ [\si{\cps}]     & $-14$  &$\pm$& $2$       & $-15$ &$\pm$& $1$       & $-13$ &$\pm$& $2$        & $-14$ &$\pm$& $2$       \\
        $\mathcal{N}$ [\si{\cps}]          & $460$  &$\pm$& $30$      & $480$ &$\pm$& $30$      & $490$ &$\pm$& $50$       & $530$ &$\pm$& $60$      \\
        $\Delta_L$ [\si{\GHz}]               & $23$   &$\pm$& $2$       & $22$  &$\pm$& $1$       & $21$  &$\pm$& $2$        & $20$  &$\pm$& $2$       \\
        $\delta\f_\mathrm{off}$ [\si{\GHz}]  & $-1.4$ &$\pm$& $0.7$     & \multicolumn{3}{c}{---} & $-2$  &$\pm$& $4$        & \multicolumn{3}{c}{---} \\  
        $\varphi_\DM$                        & \multicolumn{3}{c}{---}  & \multicolumn{3}{c}{---} & $3$   &$\pm$& $2$        & $2$   &$\pm$& $\pi$      \\
        $\delta\f_\DM$ [\si{\GHz}]           & \multicolumn{3}{c}{---}  & \multicolumn{3}{c}{---} & $5$   &$\pm$& $3$        & $5$   &$\pm$& $2$       \\
        \hline\hline
    \end{tabular}
    \caption{Best-fit parameters of the ODR and sample means from the MCMC in the setup without~(SM) and with DM with a characteristic angular frequency $\omega_\DM=\SI{1}{\afu}$. In the ODR DM fit, no detuning offset $\delta \f_{\mathrm{off}}$ was introduced since it is covered by the uncertainties on the detuning. The uncertainties on $\varphi_{\DM}$ cover the full parameter range.}
    \label{tab:bestfits}
\end{table}

\Cref{fig:Barad-Dur_vs_Minas-Tirith} depicts the data points with the uncertainties $\sigma_{\delta\f}$ on the frequency detuning $\delta\f$ and and the uncertainties $\sigma_{\dN}$ on the PMT counts \dN in red. 
The uncertainties $\sigma_{\delta\f}$ on the frequency detuning take into account the triangular modulation of the laser frequency applied during the scan.
The wings, shaded in green, correspond to the control region, in which the uncertainties $\sigma_{\dN}$ on the PMT counts were determined. 

The ODR fit directly takes into account these uncertainties, whereas the MCMC fit uses the the uncertainties~$\sigma_n$ shown in orange. These were obtained by translating the uncertainties on $\delta\f$ into uncertainties on $\dN$ and adding them in quadrature to the uncertainties on $\dN$.
The fitted excitation spectra (using ODR) with and without a DM background, with a DM oscillation frequency of $\omega_\DM = \SI{1}{\afu}$, are shown as blue and black solid curves, respectively.
The shaded bands around the lines indicate the respective $2\sigma$ uncertainty regions based on the $2\sigma$ confidence intervals~(CIs) around the best-fit values.
Both fits exhibit good agreement with the data.

For comparison, we also show the excitation spectrum resulting from a DM oscillation amplitude of $\delta\f_\DM = \SI{15}{\GHz}$, which (for sufficiently small $\Delta_L$) leads to a two-peak structure, as predicted by \cref{eq:intensity_adiabatic}.
As the lineshape deviates significantly from a Lorentzian, the corresponding spectrum is excluded by the data. However, it illustrates the expected lineshape in the DM-dominated case and highlights how the experimental data provided by Ref.~\cite{Tiedau:2024obk} can already set bounds on $\delta \f_\DM$.

\begin{figure}
    \centering
    \includegraphics[width=\columnwidth]{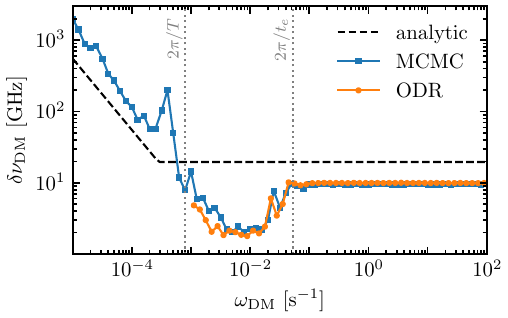}
    \caption{%
        Comparison of the analytic estimate~(black dashed) from \cref{eq:limit_slow_adiabatic,eq:limit_fast_adiabatic} of the upper bound on the DM oscillation amplitude $\delta\f_\DM$ with the \SI{95}{\%} confidence level bounds obtained from MCMC sampling~(blue squares) or ODR~(orange dots).
        To guide the eye, the numeric data points are connected by the blue and orange line, respectively.
        The vertical dotted lines indicate the DM oscillation frequencies corresponding to the excitation time~\te and the time delay~$T=\SI{130}{\minute}$ between the two scans.
    }
    \label{fig:fit_comparison}
\end{figure}

\Cref{fig:fit_comparison} depicts the bounds on $\delta\f_\DM$ as a function of $\omega_\DM$, based on an analysis of the measured lineshape presented in Ref.~\cite{Tiedau:2024obk}. Note that the bounds shown here are independent of the respective coupling causing the variation.
We show the \SI{95}{\%} upper limit derived from the MCMC sampling in blue and the corresponding $2\sigma$ bounds obtained using ODR in orange.
This allows us to put bounds on DM whilst marginalizing over the random phase $\varphi_\DM$ and consistently treating the parameter region with $\omega_\DM T\sim 2\pi$. 

The $2\sigma$ bound from our analytic estimate in \cref{eq:limit_slow_adiabatic,eq:limit_fast_adiabatic} is indicated by the dashed line in \cref{fig:fit_comparison}.
The estimate agrees within an $\mathcal{O}(1)$ factor with the numerical results in the low-frequency regime ($\omega_\DM\lesssim 2 \pi/T$, where $T=\SI{130}{\minute}$ is the time difference between the starts of the two scans).
At intermediate frequencies, $2\pi/T < \omega_\DM < 2\pi/t_e$, each excitation cycle experiences only a fraction of a DM oscillation, while the complete scan still sees at least one full oscillation.
The effect of the DM oscillations in this regime hence depends on the details of the detuning scan, and are not properly captured by our naive estimate.
The analytic estimate therefore deviates from the numerical results, while the two numerical methods agree.
In the high-frequency regime~($\omega_\DM \gtrsim 2\pi/t_e)$, the numeric bounds become constant, qualitatively agreeing with our estimate in \cref{eq:limit_fast_adiabatic}, but being roughly a factor two stronger than the estimate, as DM oscillations are not only constrained by the width of the resonance, but also by its shape (cf.\ \cref{fig:Barad-Dur_vs_Minas-Tirith} and discussion below).
Note that we did not include information on the laser parameters, but marginalized over the linewidth~$\Delta_L$ and normalization factor~$\mathcal{N}$ using generous priors.
Good prior knowledge of the laser lineshape could allow to place even stronger bounds.

%
\section{Bounds on variations of the QCD scale}
\label{sec:LambdaQCDbounds}
%

In \cref{fig:variation_QCD_scale}, we compare bounds on periodic variations of the QCD scale, which, we recall, can be probed via
\begin{align}
    \frac{\delta \f_\DM}{\f_0} = K_{\Lambda_{\mathrm{QCD}}}\,\frac{\delta \Lambda_{\mathrm{QCD}}}{\Lambda_{\mathrm{QCD}}} 
    \approx \mathcal{O}(10^5)\frac{\delta \Lambda_{\mathrm{QCD}}}{\Lambda_{\mathrm{QCD}}}\,,
\end{align}
where we assumed $K_{\Lambda_{\mathrm{QCD}}} \sim \si{\MeV}/\SI{10}{\eV} \sim 10^5$ (see also \cref{eq:Ks} and discussion in \cref{sec:Th_sensitivity}).
The lineshape bound obtained from the analysis presented in \cref{sec:lineshape_bounds} is indicated by the red line.
For comparison we show existing bounds resulting from atomic clocks~\cite{Banerjee:2023bjc,Hees:2016gop,Kennedy:2020bac,Kobayashi:2022vsf,Sherrill:2023zah} and other oscillators~\cite{Oswald:2021vtc,Campbell:2020fvq,Zhang:2022ewz} in dark and light cyan, respectively.
Due to the enhanced sensitivity of the nuclear transition, we are able to place a bound that, for modulation frequencies $f=\omega_\DM/(2\pi) \approx\SI{e-2}{\hertz}$\,, is only one to two orders of magnitude weaker than the state of the art. 
The reach of our bound is limited by the observed linewidth $\sim\SI{20}{\GHz}$ that is mainly due to the width of the laser. 
The same holds for the linewidth of $\sim\SI{300}{\kHz}$ observed by Ref.~\cite{Zhang:2024ngu}. 
We have indicated the approximate reach of this measurement, based on our analytic estimates in \cref{sec:bounds_estimate}, in orange. 
As can be seen, this measurement easily surpasses existing bounds for modulation frequencies above $f\approx\SI{e-2}{\hertz}$. 

With the development of laser systems specifically designed for the excitation of the \Th transition, the laser width is expected to further decrease to the point at which the linewidth of the transition itself will become the limiting factor. Assuming that such measurements will be carried out in solid-state systems like CaF$_2$, the random magnetic field caused by the $^{19}$F nuclei in the host material broadens the linewidth to a level of a few $\SI{100}{\hertz}$~\cite{Kazakov_2012,Rellergert:2010}. Line broadening from the second-order Doppler effect at a temperature of \SI{150}{\K} as used in the experiment \cite{Tiedau:2024obk} is in the same range.
We have indicated the reach of a measurement limited by this broadening in purple. At modulation frequencies larger than $\SI{100}{\hertz}$, sideband searches can be performed. The linewidth could be reduced below \SI{100}{\hertz}, for example, with laser-cooled \Th ions in an ion trap.

Repeated comparison of frequency standards, commonly referred to as clock comparisons, will ultimately surpass searches based on the lineshape. Indeed, in solid state hosts one is able to leverage the large number of \Th nuclei to determine the transition frequency, i.e.\ the peak of the spectrum, with much greater precision than its width, $\sigma_\f\ll\Delta \f$. This tendency is already apparent in the results of Ref.~\cite{Zhang:2024ngu}, where the reported uncertainty on the peak position is about two orders of magnitude smaller than the observed width. Assuming that for $T=\SI{e6}{\s}$, that is approximately two weeks, the \Th transition frequency is compared every $t=\SI{2}{h}$ with an uncertainty of $\sigma_\f=\SI{5}{\kHz}$, one is able to constrain oscillation amplitudes
\begin{equation}
    \delta\f_\DM\lesssim \sigma_\f \sqrt{\frac{T}{t}}
\end{equation}
over the frequency interval $2\pi/T<\omega_\DM<2\pi/t$\,. The reach of such a rudimentary clock is indicated in orange in \cref{fig:variation_QCD_scale} and should be achievable with the setup used in Ref.~\cite{Zhang:2024ngu}.

Finally, in blue we show the reach of a single ion clock limited by quantum projection noise~(QPN) with parameters typical for today's optical clocks: If operated for $T=\SI{e6}{\s}$, such a clock should allow the transition frequency to be determined with an uncertainty of $\sigma_\f=\SI{1}{\hertz}$ every $t=\SI{1}{s}$. Such experiments are in theory sensitive to arbitrarily large modulation frequencies $\omega_\DM$, although with decreased sensitivity \cite{Derevianko:2016vpm}. 
In practice, clock comparison campaigns only quote bounds for $\omega_\DM\ll 2\pi/t$, which is why the current bounds from the most precise atomic clocks (dark cyan) only reach masses up to $\sim\SI{e-16}{eV}$. 
Dedicated measurement campaigns will be required to set bounds at higher frequencies, also in the era of the
nuclear clock (blue dashed). 
Such campaigns may decrease the runtime $t=\SI{1}{s}$, which, however, is limited by experimental considerations and results in larger errors $\sigma_\f\approx 1/t$. 
Alternatively, one might make use of dynamical decoupling, which enhances the sensitivity beyond the level considered here~\cite{Aharony:2019iad} but limits the search to a very narrow frequency band for one measurement at a time.

In contrast, our method probes high frequency modulation by searching for modified lineshapes. Since these can be recorded over long timescales compared to the oscillation, our method provides bounds on frequency modulation beyond $\omega_\DM  \sim 2\pi/t$, there is no (data-sampling rate-dependent) cut-off.
Bounds from thorium lineshape analyses are therefore expected to be the dominating laboratory bound for masses $\gtrsim\SI{e-16}{eV}$, even when the first QPN limited nuclear clocks become operational.

%
\section{Interpretation in terms of dark matter models}\label{sec:models}
%

We now interpret our limits on variations of the \Th frequency in terms of two concrete models of ULDM: a scalar field with linear couplings, and the QCD axion.
Scalar ULDM generically couples linearly to the hadron masses, whereas a pseudo-scalar (axion) couples quadratically~\cite{Kim:2022ype}. However, one can construct a broad class of natural ULDM models where the leading DM interaction with the SM fields is quadratic~\cite{Banerjee:2022sqg}.
All lead to oscillations of the nuclear/hadronic parameters. Bounds on models coupling to the electromagnetic sector can be derived by recasting the bounds presented in this section.

\subsection{Scalar dark matter}

At low energies, the general interaction of a scalar $\phi$ linearly coupled to the SM is given by
\begin{align}\begin{aligned}
    \mathcal{L}_\phi &= \bigg[
    \frac{d_e}{4e^2}F_{\mu\nu}F^{\mu\nu}
    - \frac{d_g\beta_s}{2g_s}G_{\mu\nu}^A G^{A\mu\nu}
    - d_{m_e}m_e\, \bar{e}e 
    \\&\hspace{5em}
    - \sum_{q=u,d}\left(d_{m_q} 
    + \gamma_{m_q}d_g\right) m_q\bar{q}q
    \bigg] \kappa \,\phi\,,
\end{aligned}\end{align}
where $d_i$, $i=e,g,m_e,m_q$, are dimensionless couplings, $\beta_s$ is the QCD beta function and $\gamma_{m_q}$, $q=u,d$ are the anomalous dimensions of the $u$ and $d$ quarks, while $\kappa=\sqrt{4\pi}/\MPl$ denotes the inverse of the reduced Planck mass. 
The scalar couplings can in principle be related to variations of the \Th frequency, however as already discussed, at presence they are subject to significant uncertainties (see~\cite{Caputo:2024doz} for a comprehensive discussion). We shall focus on the sensitivity to variation in the strong sector, and simply assume 
\begin{align}
    \frac{\delta\f_\DM}{\f_0} \approx K_{\Lambda_{\mathrm{QCD}}} \,d_g\,\kappa\,\phi\,, 
    \label{eq:sensitivity_scalar_dm}
\end{align}
where $K_{\Lambda_{\mathrm{QCD}}}\sim \num{e5}$ (see also \cref{eq:Ks} and discussion in \cref{sec:Th_sensitivity}).

\begin{figure*}
    \includegraphics{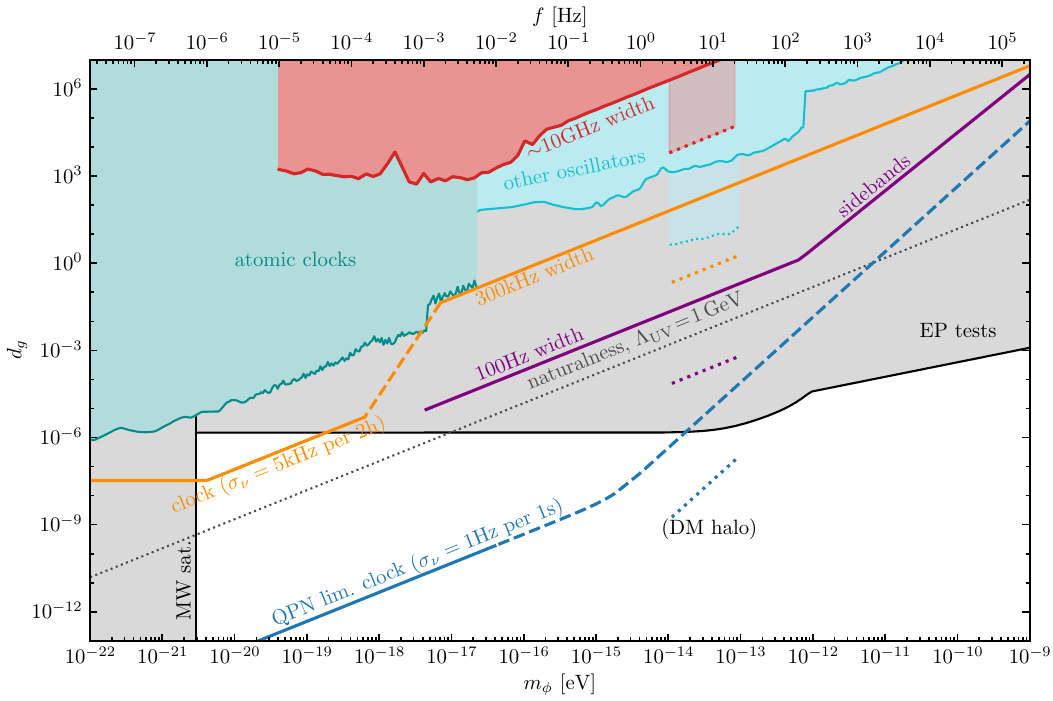}
    \caption{%
        Searches for scalar DM coupled to QCD. Colored bounds are searches for time variations of frequencies due to the scalar being DM, while the gray areas are constrained by DM being heavy enough to allow for Milky-Way~(MW) satellite galaxies~\cite{DES:2020fxi} as well as equivalence principle violating forces mediated by the scalar~\cite{Schlamminger:2007ht,Smith:1999cr,MICROSCOPE:2022doy}. 
        The lineshape bound derived in this work is shown in red, whereas the strongest current bounds~\cite{Hees:2016gop,Kennedy:2020bac,Campbell:2020fvq,Oswald:2021vtc,Kobayashi:2022vsf,Zhang:2022ewz,Sherrill:2023zah,Banerjee:2023bjc} are indicated by the light and dark cyan regions.
        Forecasts for lineshape bounds using current data with a linewidth of \SI{300}{\kHz}~\cite{Zhang:2024ngu} and for future measurements reducing the linewidth to \SI{100}{\Hz} are depicted by the orange~(right solid part of the line) and purple lines, respectively. 
        A projection for a clock operation is also shown in orange~(left solid part), and the blue line is a future QPN-limited nuclear clock~\cite{Peik:2020cwm,Arvanitaki:2014faa,Banerjee:2020kww}.
        The dotted colored lines indicate the DM-density-enhanced sensitivity in case of the presence of a DM halo around the Sun~\cite{Budker:2023sex,Gorghetto:2024vnp}.
        The parameter space below the gray dotted line is motivated by naturalness considerations, assuming a UV cutoff of \SI{1}{\GeV}.
    }
    \label{fig:scalar}
\end{figure*}

In \cref{fig:scalar} we recast the bounds in terms of the scalar's mass $m_\phi$ and its dilatonic coupling $d_g$. 
The red shaded region indicates the lineshape bound derived in this work from the excitation spectrum in Ref.~\cite{Tiedau:2024obk}, the orange line corresponds to the projection for limits using the data from Ref.~\cite{Zhang:2024ngu}. In the mass region $m_\phi < \SI{e-18}{eV}$, we already assume clock operation (continuous comparison of frequency standards) with lineshape scans repeated every two hours. The purple line is a projection of a lineshape bound in the case where the linewidth is limited by the crystal properties instead of the VUV laser.
Current clock constraints~\cite{Hees:2016gop,Kennedy:2020bac,Campbell:2020fvq,Oswald:2021vtc,Kobayashi:2022vsf,Zhang:2022ewz,Sherrill:2023zah,Banerjee:2023bjc} are shown in cyan, whereas the blue line indicates the prospective reach of a full-fledged nuclear clock~\cite{Peik:2020cwm,Arvanitaki:2014faa,Banerjee:2020kww}.

Assuming the field $\phi$ constitutes DM, its mass is bounded from below to $m_\DM \gtrsim \SI{e-21}{\eV}$ by astrophysical observations, e.g.\ of Milky-Way~(MW) satellite galaxies~\cite{DES:2020fxi} (see Ref.~\cite{Hui:2021tkt} for a review). Further, this scalar mediates equivalence-principle~(EP) violating forces that are constrained by the non-observation of relative acceleration between various test masses \cite{Schlamminger:2007ht,Smith:1999cr,MICROSCOPE:2022doy}. 
Both constraints are indicated by the gray shaded regions in \cref{fig:scalar}.
Crucially, the EP bounds are independent of the assumption that the scalar constitutes DM, for which reason their relative strength with respect to the bounds on the time-variation of fundamental constants depends on the local DM density~\cite{Banerjee:2019xuy}. In the recasting of the bounds, we assumed that the scalar constitutes all of the DM and that the DM density coincides with the one inferred from Milky Way observations, i.e.\ $\rho_\DM=\SI{0.4}{\GeV/\cm^3}$~\cite{McMillan:2011wd}. For the masses $m_\phi=\left(\num{e-14}\text{~--~}\num{e-13}\right)\,\si{\eV}$, we further indicate by dotted lines how the existence of a DM halo with a relative over-density of $10^5$ around the Sun enhances the reach of the searches discussed here. Such a halo might form via the mechanism presented in Ref.~\cite{Budker:2023sex,Gorghetto:2024vnp}. 

For scalar masses in the range of \SI{5e-17}{\eV} to \SI{5e-16}{\eV}, the bound derived in this work~(red) falls shy of the current leading clock bounds~(cyan) by only one to two orders of magnitude.
Our projection for a similar analysis using currently existing data~\cite{Zhang:2024ngu} outperforms the clock bounds above $m_\phi \gtrsim \SI{e-17}{\eV}$~(right solid part of the orange line).

While the lineshape bounds cannot reach beyond the exclusion limits of EP tests, even when pushing the linewidth measurement to the \SI{100}{\Hz} limit, a clock operation of the existing laser-excitation setup~(left solid part of the orange line) can access unprobed parameter space in the mass range $m_\phi \simeq \SIrange{3e-21}{2e-19}{\eV}$.
A nuclear clock operated at the QPN limit surpasses the EP tests below masses of $m_\phi \lesssim \SI{e-14}{\eV}$ and can even probe the parameter region motivated by naturalness, i.e.\ where the quantum loop corrections to the scalar mass are on the order of or less than the mass itself, $\Delta m_\phi^2 \sim d_g^2\,\Lambda_\mathrm{UV}^4/(4 \pi M_\mathrm{Pl})^2 \lesssim m_\phi^2$, assuming a UV cutoff of $\Lambda_\mathrm{UV}\sim\SI{1}{\GeV}$, indicated by the dotted line.

%
\subsection{Axion}
%

As an example of a quadratically coupled field, we consider the QCD axion~$a$. It arises as the pseudo-Goldstone boson in the spontaneous breaking of the Peccei-Quinn symmetry and exhibits the defining coupling
\begin{align}
    \mathcal{L}_a\supset\frac{g_{s}^2}{32\pi^2}\frac{a}{f_a}G_{\mu\nu}^A \tilde{G}^{A\mu\nu}\,,
\end{align}
where $\tilde{G}$ is the dual of the gluon field strength. Once QCD confines, this coupling gives rise to interactions between pions and the axion, inducing an axion-dependence of the pion mass.
\begin{align}
    m_\pi^2(\theta)=B(m_u^2+m_d^2+2m_u m_d \cos(\theta))^{1/2} \label{eq:chiral_pion_mass}
\end{align} 
with $\theta=a/f_a$ and $B=\langle\overline q q\rangle/f_\pi^2$. Expanding around $\theta=0$, one finds the leading quadratic interaction $\propto \theta^2\pi^2$. 
Note that these effects are due to the QCD anomaly and for a generic axion-like particle suppressed by $m_a^2/f_a^2$ \cite{Banerjee:2022sqg}. 
Following \cite{Kim:2022ype}, we find for the QCD axion
\begin{equation}
    \frac{\delta \f_{\DM}}{\f_{0}} = -\num{2.2e4}\  \theta^2\,.
    \label{eq:sensitivity_axion}
\end{equation}

\begin{figure*}
    \includegraphics{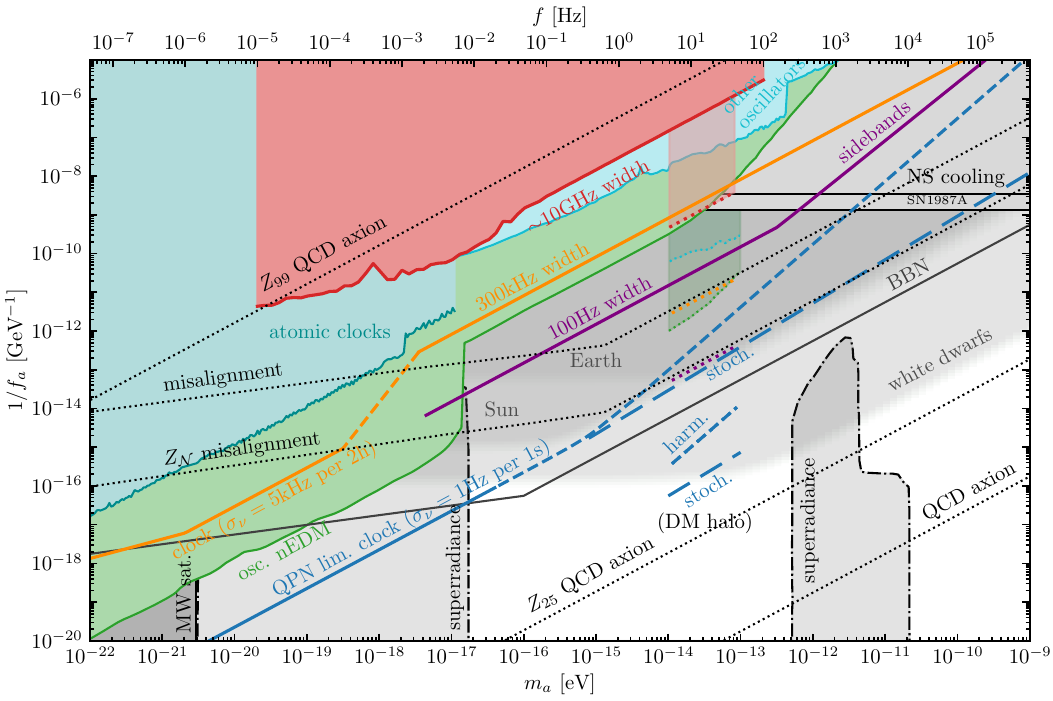}
    \caption{%
        Searches for axion DM. 
        The lineshape bound derivated in this work is shown in red, whereas the strongest current bounds~\cite{Hees:2016gop,Kennedy:2020bac,Campbell:2020fvq,Oswald:2021vtc,Kobayashi:2022vsf,Zhang:2022ewz,Sherrill:2023zah,Banerjee:2023bjc} and limits from oscillating nEDM searches~\cite{Abel:2017rtm,Schulthess:2022pbp} are indicated by the cyan and green regions, respectively.
        Forecasts for lineshape bounds using current data with a linewidth of \SI{300}{\kHz}~\cite{Zhang:2024ngu} and for future measurements reducing the linewidth to \SI{100}{\Hz} are depicted by the orange~(right solid part of the line) and purple lines, respectively. 
        A projection for a clock operation is also shown in orange~(left solid part), and the blue line is a future QPN-limited nuclear clock~\cite{Peik:2020cwm,Arvanitaki:2014faa,Banerjee:2020kww}.
        In the mass region $\SI{e-14}{\eV} \lesssim m_a \lesssim \SI{e-13}{\eV}$, we further show the respective extended reach in the case of the presence of an $\mathcal{O}\num{e5}$ DM overdensity around the Sun~\cite{Budker:2023sex,Gorghetto:2024vnp} as dotted colored lines.
        Constraints independent of the axion abundance are shown in gray. 
        These include the lower bound on the DM mass from Milky-Way satellite galaxies~\cite{DES:2020fxi} (MW sat., dark gray), lower bounds on the axion decay constant from cooling of neutron stars~(NS)~\cite{Leinson:2021ety} and SN1987A~\cite{Raffelt:2006cw,Carenza:2019pxu,Caputo:2024oqc}, and superradiance constraints~\cite{Unal:2020jiy,Arvanitaki:2014wva}~(dash-dotted contours).
        In the parameter space above the dark gray solid line, the axion coupling would alter predictions of BBN~\cite{Blum:2014vsa}, whereas in the gray shaded areas, the axion is constrained due to density corrections to its potential inside Earth, Sun or white dwarfs~\cite{Hook:2017psm,Balkin:2022qer,Balkin:2023xtr}, respectively. 
        The black dotted lines labeled '($Z_\Ncopy$) QCD axion' depict the decay constant as a function of the axion mass predicted in the standard QCD axion scenario or for light QCD axions in models with $\Ncopy$ copies of QCD~\cite{Hook:2018jle}, whereas the black dotted lines labeled '($Z_\Ncopy$) misalignment' indicate the maximal value of $1/f_a$ for which the DM abundance can be obtained in the standard misalignment scenario~\cite{Preskill:1982cy,Abbott:1982af,Dine:1982ah}.
    }
    \label{fig:axion}
\end{figure*}

In \cref{fig:axion} the corresponding bounds are shown in terms of the axion mass $m_a$ and the inverse decay constant $1/f_a$\,. 
Currently, the best laboratory bounds on axion DM come from searches for an oscillating neutron electric dipole moment (nEDM), shown by the green shaded area \cite{Abel:2017rtm,Schulthess:2022pbp}. 
We further indicate in gray the parameter space excluded by the non-observation of superradiance~\cite{Unal:2020jiy,Arvanitaki:2014wva} or by supernova~(SN)~\cite{Raffelt:2006cw,Carenza:2019pxu,Caputo:2024oqc} and neutron star~(NS) cooling bounds~\cite{Leinson:2021ety}, big bang nucleosynthesis~(BBN)~\cite{Blum:2014vsa}, as well as bounds on the field being sourced by massive objects like the Earth, Sun or white dwarfs \cite{Hook:2017psm,Balkin:2022qer,Balkin:2023xtr}.

In a minimal model of the QCD axion, \cref{eq:chiral_pion_mass} relates the axion mass to the decay constant via the QCD scale $m_a\sim \Lambda_\mathrm{QCD}^2/f_a$ (dotted line labeled \emph{QCD axion} in \cref{fig:axion}). In this work we are, however, studying a broader class of models in which the QCD axion can be lighter either because its mass is fine-tuned or as the result of a large discrete $Z_\Ncopy$ symmetry \cite{Hook:2018jle} (dotted lines labeled \emph{$Z_{25/99}$ QCD axion} in \cref{fig:axion}). In \cref{fig:axion}, we indicate with dotted lines the $1/f_a$ values below which the observed DM abundance may be obtained via the minimal misalignment mechanism~\cite{Preskill:1982cy,Abbott:1982af,Dine:1982ah} in the fine-tuned and $Z_\Ncopy$ symmetry cases, respectively. We assumed that reheating happens shortly before BBN $T_\mathrm{rh} \gtrsim T_\mathrm{BBN} \sim \SI{1}{\MeV}$\,, which gives the largest DM densities.

For the axion, the lineshape analysis carried out in this work~(red) yields bounds that are competitive with current clock bounds~(cyan) around $m_a\sim\SI{2e-17}{\eV}$ and only about one order of magnitude weaker than the clock bounds at higher masses.
Our projection for the data of Ref.~\cite{Zhang:2024ngu}~(orange) surpasses current clocks and is only an order one factor less sensitive than oscillating nEDM searches. 
When reaching a linewidth of $\sim\SI{100}{\Hz}$~(purple), lineshape bounds are about one order of magnitude more sensitive than the nEDM above $m_a \gtrsim \SI{e-17}{\eV}$, and a full-fledged nuclear clock~(blue solid line) will overcome the nEDM bound even below $m_a \lesssim \SI{e-17}{\eV}$ by roughly two orders of magnitude.

DM candidates that are quadratically coupled to the SM and have masses larger than \SI{e-15}{\eV}, can, besides the harmonic variation of transition frequencies discussed so far, also give rise to stochastic variations over the experiment's time scales \cite{Masia-Roig:2022net,Flambaum:2023bnw,Kim:2021yyo,Kim:2023pvt}. 
These fluctuations are due to the DM velocity dispersion $v_\DM$ and feature a white noise spectrum with a UV cut-off at $\f_\mathrm{mod}\sim v_\DM^2m_a$\,.
The velocity dispersion within the Milky Way is estimated as $v_\DM\sim10^{-3}$ and can, within a solar halo, be assumed to be similar to the escape velocity $v_\mathrm{esc}\sim10^{-4}$. 
Such white noise can be searched for in the timing data of clock comparison experiments \cite{Kim:2023pvt}. 
In \cref{fig:axion} we indicate with a blue long-dashed line the corresponding reach of a QPN-limited clock.

%
\section{Summary}
\label{sec:conclusion}
%

In this work, we discuss the implications of the laser excitation of \Th for ultralight dark matter~(ULDM) searches. We argue that already at this early stage, before the realization of a nuclear clock, the excitation spectrum of the isomeric transition is sensitive to ULDM. The oscillation of the nuclear parameters induced by the ULDM modifies the lineshape of the transition. Using the data taken by \citeauthor{Tiedau:2024obk}~\cite{Tiedau:2024obk} we perform, for the first time, a new-physics motivated analysis of the spectrum and provide a bound on the ULDM coupling versus its mass in two well-motivated (pseudo)scalar ULDM models, as well as a model-independent assessment of the sensitivity to a variation of the QCD scale.
We also demonstrate how our method can be applied to future measurements, we provide projections for other existing laser excitations of \Th, and derive analytic estimations for the regions of slow and fast oscillating DM. 

\acknowledgements
The work of GP is supported by grants from the United States-Israel Binational Science Foundation~(BSF) and the United States National Science Foundation~(NSF), the Friedrich Wilhelm Bessel research award of the Alexander von Humboldt Foundation, and the Israel Science Foundation~(ISF). 
EF and FK acknowledge funding by the Deutsche Forschungsgemeinschaft (DFG, German Research Foundation) under Germany’s Excellence Strategy -- EXC-2123 QuantumFrontiers -- 390837967 -- and support by the SFB 1227 (DQ-mat) -- Project-ID 274200144.
The work of EP and JT has been funded by the European Research Council (ERC) under the European Union’s Horizon 2020 research and innovation programme (Grant Agreement No. 856415), the Deutsche Forschungsgemeinschaft (DFG) -- SFB 1227 -- Project-ID 274200144 (Project B04), and by the Max-Planck-RIKEN-PTB-Center for Time, Constants and Fundamental Symmetries.

\appendix

\section{Emergence of sidebands}
\label{sec:sidebands}

The analysis presented in \cref{sec:bounds_estimate} applies for $\delta\f_\DM \gg \omega_\DM/(2\pi)$. Here we derive the phenomenology for cases where $\delta \f_\DM \ll \omega_\DM/(2\pi)$ and rather than a single, broadened peak, multiple sidebands can be observed.

\begin{figure}
    \centering
    \includegraphics[width=\columnwidth]{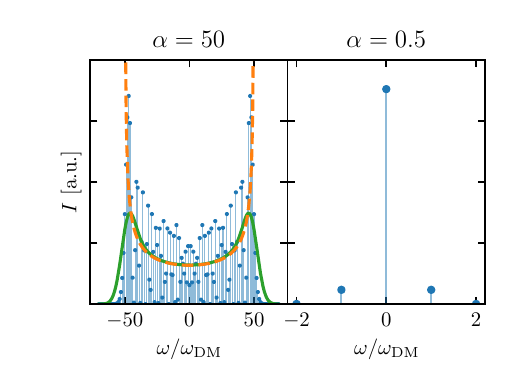}
    \caption{DM induced sidebands at large~(left) and small~(right) modulation index $\alpha$.}
    \label{fig:sidebands}
\end{figure}

Sidebands arise whenever a carrier signal (in this case at the nuclear transition frequency \fTh) is modulated by a signal of a different frequency (in this case, the characteristic frequency $\omega_{\DM}/(2\pi)$ of the oscillating DM).
For illustration, let us consider a two-level system, where, without loss of generality, we take the energy of the ground state as 0, while the energy of the excited state is given by \cref{eq:oscillating freq} and we set $\varphi_\DM=0$. The excited state~$\left|\uparrow\right>$ evolves as 
\begin{align}\begin{aligned}
    \left|\psi_{\uparrow}(t)\right>&=\exp\left[-i2\pi\f_0 t-i\alpha\sin(\omega_\DM t)\right]\left|\uparrow\right>\\
    &=\sum_{n=-\infty}^{\infty}\! J_n\!\left(\alpha\right) \exp\left[-i (2\pi\f_0+n \omega_\DM) t\right]\left|\uparrow\right>,
\end{aligned}\end{align}
where we used the Jacobi-Anger expansion in the second step, $J_n$ denotes the $n$-th Bessel function and $\alpha =2\pi\delta\f_\DM/\omega_\DM$ is the modulation index.
The result suggests that the transition can be resonantly driven at frequencies $\f_0+n \omega_\DM/(2\pi)$, although the relative rate will be suppressed by the statistical weight $|J_{n}(\alpha)|^2$. Therefore, the lineshape is convoluted with
\begin{equation}
    I(\f)=\sum_{n=-\infty}^{\infty} \left|J_n \left(\alpha\right)\right|^2\delta\left[\f-\left(\f_0+n\,\frac{\omega_\DM}{2\pi}\right)\right]
    \label{eq:intensity_sidebands}
\end{equation}
instead of \cref{eq:intensity_adiabatic}. It can be seen from the right side of \cref{fig:sidebands} that for $\alpha\ll1$ there is a main peak at $\f_0$, as $|J_0(\alpha)|^2\approx 1$, together with two sidebands at $2\pi \f_0\pm\omega_\DM$, with an intensity suppressed by $|J_1(\alpha)|^2\approx \alpha^2/4$. Peaks of higher order are even further suppressed. If an experiment is able to detect sidebands with a relative intensity $\Delta I/I$ compared to the main peak, one is therefore able to constrain
\begin{equation}
    \delta\f_\DM\lesssim \frac{\omega_\DM}{\pi} \sqrt{\frac{\Delta I}{I}}\,.
\end{equation}
Note that in the opposite limit of $\alpha\gg1$, all sidebands up to $n\simeq \alpha$ contribute significantly. In fact, it can be shown that in this limit \cref{eq:intensity_adiabatic,eq:intensity_sidebands} for $I(\f)$  are equivalent if the laser linewidth is large compared to $\omega_\DM$. This is depicted in the left side of \cref{fig:sidebands}, where we show, for $\alpha=50$\,, \cref{eq:intensity_sidebands} convoluted with a Gaussian of width $5\,\omega_\DM$ in green, as well as \cref{eq:intensity_adiabatic} in orange.

\section{Details of the numerical analysis}
\label{sec:MCMCandODR}

We here provide further details on our analysis of the \Th fluorescence signal recorded in Ref.~\cite{Tiedau:2024obk}, as described in \cref{sec:Th_excitation}.
Denoting the number of data points in the frequency scans (shown in red in \cref{fig:Barad-Dur_vs_Minas-Tirith}) by $N_{\text{points}}$, we define $t_n=t_0 + n \,(\te+\td)$ and $T_n = t_n + t_e$, $n=0, \ldots, N_{\text{points}}$, such that the excitation periods are $[t_n, T_n]$, whereas the detection periods are $[T_n, t_{n+1}]$. 
Defining the average count rate during each  excitation and detection cycle as $N_n$, we model the post-processed count rate by
\begin{align}
    \dN_n \equiv &\left(N_n - N_{\mathrm{bkg}}\right) - e^{-\frac{t_e + t_d}{\tau}} \left(N_{n-1} - N_{\mathrm{bkg}}\right),
    \label{eq:post_processing}
\end{align}
where we subtracted the constant background~$N_{\mathrm{bkg}}$ due to the radioactivity of the crystal. The second term removes fluorescence photons detected in a given cycle~$n$ but excited during previous cycles.

The number of nuclei excited between times $t$ and $t+dt$ is given by
$d N_e = \Gamma(t) d t$,
where the excitation rate 
$\Gamma(t)$ is proportional to the convolution of the laser profile $L(\f;\f_L)$ and the \Th nuclear resonance lineshape $I(\f;\fTh)$ inside the crystal,
\begin{align}
    \Gamma(t) \propto \int\! d\f\, L\big(\f;\f_L(t)\big)\, I\big(\f;\fTh(t)\big)\,,
\end{align}
with $\f_L$ and $\fTh$ denoting the laser peak frequency and \Th resonance, respectively.
The probability of a nucleus excited at time $t'$ to decay between $t$ and $t + dt$ is 
$p(t,t') dt = \exp\left({-\frac{t-t'}{\tau}}\right) \frac{dt}{\tau}$.
The PMT count rate~$\dN_n$ from fluorescence photons produced from nuclei excited between $t_n$ and $T_n$ and recorded between $T_n$ and $T_n+\td$ is hence given by
\begin{align}\begin{aligned}
    \dN_n
    &= \varepsilon \int\limits_{T_n}^{T_n+\td} \hspace{-.75em}dt \int\limits_{t_n}^{T_n} \!dt' \, \Gamma(t')\,p(t,t') \\
    &= \varepsilon\,\frac{1}{\td} \left(1-e^{-\frac{\td}{\tau}}\right) \int\limits_{t_n}^{T_n}\! dt\ \Gamma(t)\,e^{\frac{t-T_n}{\tau}}\,,
\end{aligned}\end{align}
where $\varepsilon$ is the efficiency for converting fluorescence photons into PMT counts.

We assume the laser frequency to be constant within each excitation interval.
The time-dependence of the excitation rate is then dominated by that of the detuning $\delta \f = \f_L-\fTh$ from the \Th resonance.
Approximating the laser profile as a Lorentzian with full linewidth at half maximum~$\Delta_L$ and the \Th spectrum by a Dirac $\delta$-distribution, we obtain
$\Gamma(t) \propto \left[1 + 4\, (\delta\f(t)/\Delta_L)^2\right]^{-1}$.
Substituting $x = (T_n-t)/\te$ and absorbing all constant prefactors into the normalization factor $\mathcal{N}$, we obtain \cref{eq:pp_count_rate},
where we include an additional count offset~$\dN_\mathrm{off}$.
The detuning~$\delta\f_n$ of the laser from the \Th resonance during the $n$-th measurement consists of the DM-independent detuning~$\delta\f_{0,n}$, a constant offset~$\delta\f_\mathrm{off}$ due to the unknown \Th frequency, and the DM-induced frequency modulation~$\delta\f_\DM$, i.e.\ 
$\delta\f_n(x) =  \delta\f_{0,n} + \delta\f_\mathrm{off} + \delta\f_\DM \cos(\omega_\DM\te \,x - \varphi_n)$, where $\varphi_n$ is the DM phase at the beginning of each measurement period.
If the DM oscillations are coherent throughout the scan, the phase in each bin is given by 
$\varphi_n = \varphi_\DM + n\,\omega_\DM\,(\te + \td)$.

For the scan in Fig.~2 of Ref.~\cite{Tiedau:2024obk}, we have $\te = \SI{120}{\s}$, $\td = \SI{150}{\s}$, and $\tau=\SI{630}{\s}$. 
The total duration of the experimental run is roughly four hours, including a scan decreasing the laser frequency, a $\sim\SI{35}{\min}$ break and then a scan increasing the frequency.
Hence, DM oscillations with angular frequencies below roughly $\omega_\DM\lesssim\SI{100}{\afu}$ are coherent throughout the entire scan (recall that the coherence time is $\tau_\mathrm{coh} \sim 1/(m_\DM v_\DM^2)$).
For fixed $\omega_\DM$, we can then constrain $\delta\f_\DM$ marginalizing over $\mathcal{N}$, $\Delta_L$, $\dN_\mathrm{off}$, $\delta\f_\mathrm{off}$, and~$\varphi_\DM$.

The uncertainties on the counts are modeled by the standard deviation of the counts recorded in the control regions (wings, shaded in green in \cref{fig:Barad-Dur_vs_Minas-Tirith}), which we defined as the regions corresponding to a frequency detuning $\left|\delta\f_0\right| > \SI{50}{\GHz}$. We obtain $\sigma_{\dN} \approx \SI{5}{\cps}$. 
During data taking, a $\sim\SI{5}{\GHz}$ triangular frequency modulation was applied to the VUV laser frequency.
However, as we do not know the exact amplitude and phases of the modulation, we do not include this modulation in $\delta\f_n(x)$, but instead model it by assigning a corresponding uncertainty on the detuning steps.
The uncertainties on the frequency detuning are assumed to be of the order of $\sigma_{\delta\f} \sim \SI{10}{\GHz} / \sqrt{12}\approx \SI{3}{\GHz}$, i.e.\ uniformly distributed in a $\pm \SI{5}{\GHz}$ window around the central frequency.

\subsection{Monte Carlo sampling}

Our log-likelihood for the MCMC-based curve fit is given by
\begin{align}
    \log\mathcal{L} = -\frac{1}{2} \sum\limits_n \left(\frac{\dN_n^\text{pred} - \dN_n^\text{meas}}{\sigma_n}\right)^2\,,
\end{align}
where $\dN_n^\text{meas}$ are the postprocessed PMT counts [cf.\ \cref{eq:post_processing}] 
measured in the experiment, and $\dN_n^\text{pred}$ is the respective prediction from \cref{eq:pp_count_rate}.
For the uncertainty~$\sigma_n$, we take the variance~$\sigma_{\dN}$ of the data points in the wings. 
In the region around the peak, $|\delta\f_0|<\SI{50}{\GHz}$, we further propagate the uncertainty~$\sigma_{\delta\f}$ on the detuning to the count rate and add it in quadrature,
i.e.\ $\sigma_n^2 = \sigma_{\dN}^2 + \sigma_{\delta\f}^2 \left|\frac{d y}{d \delta\f_0}(\delta\f_{0,n})\right|^2$,
where we evaluate the absolute value of the derivative as the mean of the absolute values of the slopes between each point and its left and right neighbor (orange error bars in \cref{fig:Barad-Dur_vs_Minas-Tirith}).

\begin{table}
    \vspace*{1pt}
    \centering
    \renewcommand{\arraystretch}{1.25}
    \setlength{\tabcolsep}{.5em}
    \begin{tabular}{cc|cc}
        \hline\hline
        parameter & prior range   & parameter & prior range \\\hline
        $\dN_{\mathrm{off}}$ [\si{\cps}]    & $(-100, 100)$ &
        $\mathcal{N}$ [\si{\cps}]         & $(100, 1000)$ \\
        $\Delta_L$ [\si{\GHz}]              & $(1, 100)$ &
        $\delta\f_\mathrm{off}$ [\si{\GHz}] & $(-100, 100)$ \\  
        $\varphi_\DM$                       & $(0, 2\pi)$ &
        $\delta\f_\DM$ [\si{\GHz}]          & $(0, \num{e6})$\\
        \hline\hline
    \end{tabular}
    \caption{Uniform prior ranges for the sampled parameters in the MCMC.}
    \label{tab:priors}
\end{table}

We sample the posterior distribution using the parallel tempering MCMC sampler \texttt{PTMCMCsampler}~\cite{justin_ellis_2017_1037579}.
The priors for the sampled parameters are uniform in the ranges indicated in \cref{tab:priors}.
The upper limit on the  DM oscillation amplitude~$\delta\f_\DM$, marginalized over the remaining parameters, is then obtained as the \SI{95}{\%} quantile of the MCMC samples.
The resulting bounds are shown as the blue line in \cref{fig:fit_comparison}.

\subsection{Lineshape fit using orthogonal distance regression}

Assigning the same uncertainties $(\sigma_{\delta\f}, \sigma_\dN)$ to all data points $(\delta\f_{0,n}, \dN_n^\text{meas})$ (red error bars in \cref{fig:Barad-Dur_vs_Minas-Tirith}), we perform an orthogonal distance regression on the data points, to the curve defined by $(\delta\f_0, \dN)$, with $\dN$ as defined in \cref{eq:pp_count_rate}.
In absence of DM, the fit parameters are the count offset $\dN_{\mathrm{off}}$, the normalisation factor $\mathcal{N}$, and the laser linewidth $\Delta_L$. In presence of DM with a characteristic frequency of $\omega_\DM$, the additional fit parameters are $\delta\f_\DM$ and $\varphi_\DM$. 
The detuning offset $\delta\f_\mathrm{off}$ is not fitted since it is covered by the  uncertainty~$\sigma_{\delta\f}$ on the detuning in the frequency range of interest.
We focus on the DM frequency range $\omega_\DM \gtrsim \SI{e-3}{\afu}$, since at lower frequencies the characteristic signature of DM is a slow oscillation of the full resonance, which cannot be captured by a curve fit.

The effect of the DM oscillation on the nuclear lineshape can be studied by propagating the uncertainties on the fit parameters to \dN. In \cref{fig:Barad-Dur_vs_Minas-Tirith}, we show the $2\sigma$ region for the model without DM in blue and the $2\sigma$ region for the model with DM with a characteristic frequency $\omega_{\DM}=\SI{1}{\hertz}$ in black. 
We estimate the bound on $\delta\f_\DM$ by computing $\delta\f_\DM\vert _{\text{best-fit}} + 2 \,\sigma_{\delta\f_\DM}$.
The resulting upper limits as a function of $\omega_\DM$ are indicated by the orange line in \cref{fig:fit_comparison}.
For $\omega_\DM\in \left[ 10^{-1}, 10^2\right]\,\si{\afu}$, we observe a flat frequency behaviour with a median of $\delta\f_\DM\approx \SI{10}{\GHz}$.

\bibliography{bibliography}

\end{document}